\documentclass[journal, twoside]{IEEEtran}

\usepackage{amsmath,amsfonts}
\usepackage{algorithmic}
\usepackage{algorithm}
\usepackage{array}
\usepackage[caption=false,font=normalsize,labelfont=sf,textfont=sf]{subfig}
\usepackage{textcomp}
\usepackage{stfloats}
\usepackage{url}
\usepackage{verbatim}
\usepackage{graphicx}
\usepackage{cite}
\usepackage[switch]{lineno}
\usepackage{color}
\newcommand\nc\newcommand
\nc\be{\begin{equation}}
\nc\ee{\end{equation}}
\def\abs#1{\left\vert #1 \right\vert} 
\def\mc#1{\left\vert #1 \right\vert^{2}} 
\def\cali#1{\cal#1\mit}
\nc\RR{\mathbb{R}}
\nc\vect[1]{\boldsymbol{#1}}
\nc\mean[1]{\overline{\boldsymbol{#1}}}

\newcommand{\expo}{\mbox{$\rm{e}$}}
\nc\dir[1]{\boldsymbol{\widehat #1}}
\nc\scal[2]{\boldsymbol{#1}\cdot \boldsymbol{#2}}
\newcommand\demi{\frac{1}{2}}
\nc\sign[1]{\mbox{sign}(#1)}
\newcommand\col[1]{{\color{black}#1}} 

\begin{document}
\title{Improved calculation of the second-order ocean Doppler spectrum for sea state inversion}
\author{Charles-Antoine Gu\'erin ~\IEEEmembership{(MIO, Univ Toulon, Aix-Marseille Univ, CNRS, IRD, Toulon, France)} }




\maketitle

\begin{abstract}
  We {describe and exploit a reformulation, based on a \col{recently-introduced} change of variables, of the double integral that describes the second-order ocean Doppler spectrum measured by High-Frequency radars. We show that this alternative expression, which was primarily designed for improving the numerical inversion of the ocean wave spectrum, is also advantageous for the analytical inversion of the main sea state parameters. To this end,} we revisit Barrick's method for the  estimation of the significant wave height and the mean period from the ocean Doppler spectrum. On the basis of numerical simulations we show that a better estimation of these parameters can be achieved {which} necessitates a preliminary bias correction that depends only on the radar frequency. A second consequence of {this improved} formulation is the derivation of a simple yet analytical nonlinear approximation of the second-order ocean Doppler spectrum when the Doppler frequency is larger than the Bragg frequency. This opens {up} new perspectives  for the inversion of directional wave spectra from High-Frequency radar measurements.
\end{abstract}


\section{Introduction}
High-Frequency radars have been used for half a century as an efficient tool for measuring surface currents and waves in the coastal region (see e.g. \cite{lorente_OceanScience22} for a recent review). In this range of radio frequencies (3-30 MHz), the backscattered Doppler spectrum from the sea surface is accurately described by the second-order electromagnetic and hydrodynamic perturbation theory, {the complete equations for which} were published by Barrick half a century ago \cite{barrick_tropo_1972}. Hasselmann \cite{hasselmann_Nature71} first suggested that the continuous component of the ocean Doppler spectrum is essentially a replica of the ocean wave spectrum translated by the Bragg frequency and thus could be used for sea state inversion. As many {researchers} have shown, this is only a coarse approximation and estimating the wavenumber spectrum has both mathematical and practical limitations (e.g. \cite{wyatt_JAOT00}). There is no unique relationship between the ocean spectrum and the Doppler spectrum, so that the inversion can only be performed with some additional constraints. The numerical inversion of the Doppler spectrum, using either Barrick's linearized equations \cite{lipa_RS77,wyatt_IntJRemSens90} or their fully nonlinear expression \cite{hisaki_RS96}, is computationally intensive. It requires solving a large number of direct problems involving the tricky calculation of a double integral with singular kernel and constrained variables.
{In a recent series of papers \cite{shahidi_OCEANS16,shahidi_OCEANS17,shahidi_JOE20}, it was found that an alternative change of variables can bring the integral equation relating the Doppler spectrum to the wavenumber spectrum into a form that is more adapted to the fully nonlinear numerical inversion of the latter. In an attempt to improve the accuracy and numerical efficiency of the calculation of the second-order ocean Doppler spectrum, we realized independently (before the reviewer drew our attention to the aformentioned works)  than this simple yet nonstandard change of variables is more appropriate for the evaluation of the double integral that defines the second-order Doppler spectrum (section \ref{sec:calculintegrale}). As we will see, this improved formulation is not only adapted to the numerical inversion of the wave spectrum but also unveils some useful analytical approximations.
  We revisit} the classical problem of estimating the main sea state parameters from the integral properties of the Doppler spectrum, following the method first proposed by Barrick \cite{barrick_RSE77,barrick_RS77}. To do this, we introduce a function, referred to as the ``Zeta'' function, which has the same structure as the second-order Doppler spectrum but is free of the coupling coefficient. We show that the Zeta function can, in principle, be used for the estimation of the significant wave height (SWH) and the mean period. However, numerical simulations using a typical ocean wave spectrum model show that the estimation of these two parameters must be corrected by a frequency-dependent multiplicative and additive bias, respectively (section \ref{sec:zetafunction}). Transferring these results to the actual Doppler spectrum requires the use of a weighting function, as first proposed by Barrick, to account for the influence of the coupling coefficient. In the light of this improved formulation we derive a new weighting function (Section \ref{sec:wf}) to approximate the Zeta function, resulting in a more accurate estimation of the sea state parameters. Our last result concerns the direct estimation of the ocean wave spectrum. For Doppler frequencies higher than the Bragg frequency we obtain a simple analytical approximation which is verified to be extremely accurate (Section \ref{sec:directinversion}).  For technical reasons we limit this analysis to the backscattering configuration and to the deep-water gravity wave dispersion relation for the hydrodynamic processes at play, a double assumption which corresponds to the majority of practical situations. {Note, however, that the improved change of variables that is used for the second-order Doppler spectrum can be extended to the bistatic case with little adaptation \cite{silva_OCEANS17}.}

\section{Evaluation of the second-order spectrum}\label{sec:calculintegrale}
We consider the classical idealized problem of a High-Frequency radar measuring the backscattered ocean Doppler spectrum in vertical polarization at grazing incidence in the limit of an infinite sea surface patch. We denote $f_0$ the radar frequency, $k_0=2\pi f_0/c_0$ the associated electromagnetic wavenumber, where $c_0=3\ 10^8$ m/s is the speed of light in vacuum, and $\vect k_0$ the horizontal electromagnetic wave vector of the incident electric field. As is well known in ocean radar remote sensing, the so-called { Bragg wave vector} $\vect k_B=-2\vect k_0$, { Bragg wavenumber} $k_B=2k_0$ and { Bragg frequency} $\omega_B=\sqrt{gk_B}$ play a prominent role in the expression of the backscattered field. The complete expression {for} the ocean Doppler cross-section appeared for the first time in \cite{barrick_tropo_1972}. It is given by a sum of two terms, $\sigma(\omega)=\sigma_1(\omega)+\sigma_2(\omega)$, where $\sigma_1$ contains the first-order Bragg peaks
\be\label{eq:first_order}
\sigma_1(\omega)={\cali N}\sum_{n_1=\pm 1} S_d(n_1\vect k_B)\delta(\omega-n_1\omega_B)
\ee
and $\sigma_2$ is a second-order, continuous component
\be\label{eq:second_order}
\begin{split}
  \sigma_2(\omega)&={\cali N} \sum_{n_1=\pm 1}\sum_{n_2=\pm 1} \int S_d(n_1\vect k_1)S_d(n_2\vect k_2) \abs{\Gamma}^2\\
  &\delta(\omega-n_1\omega_1-n_2\omega_2)d\vect k_1
  \end{split}
\ee
where ${\cali N}=2^6\pi K_0^4$ is a normalization constant and $S_d(\vect k)$ the directional ocean spectrum. The wave vectors entering in this integral are related by the Bragg resonance condition and the associated frequencies are given by the deep water gravity waves dispersion relation
\be\label{defk2}
\vect k_1+\vect k_2=\vect k_B,\ \omega_1=\sqrt{gk_1},\ \omega_2=\sqrt{gk_2}
\ee
where $g=9.81\ ms^{-2}$ is the gravitational constant. The kernel $\Gamma$ can be decomposed into a hydrodynamic kernel $\Gamma_H$ and an EM kernel $\Gamma_{EM}$
\be
\Gamma(\vect k_1,\vect k_2,\omega,\omega_B)=\Gamma_H+\Gamma_{EM}
\ee
with
\be
\Gamma_H=-\frac{i}{2}\left( k_1+ k_2-\frac{ ( k_1 k_2-\scal{ k_1}{ k_2})({\omega}^2+\omega_B^2)}{n_1n_2\sqrt{ k_1 k_2}({\omega}^2-\omega_B^2)}\right)
\ee
and
\be\label{eq:EMkernel}
\Gamma_{EM}=\demi \frac{(\scal{ k_1}{ k_B})(\scal{ k_2}{ k_B})-2\scal{ k_1}{ k_2}}{\sqrt{\scal{ k_1}{ k_2}}+\Delta}
\ee
where $\Delta=0.011-0.012i$ is an impedance term that takes into account the finite conductivity of the sea surface in the High-Frequency domain. Each of the four components in (\ref{eq:second_order}) corresponding to the different sign combinations $(n_1,n_2)$ addresses a different frequency domain of the Doppler spectrum. The combined geometrical constraints $\omega=n_1\omega_1+n_2\omega_2$ and $\vect k_1+\vect k_2=\vect k_B$ impose $n_1=n_2=1$ if $\omega\geq\omega_B$, $n_1=n_2=-1$ if $\omega\leq -\omega_B$ and $n_1=-n_2=\pm 1$ if $\abs\omega<\omega_B$. The different combinations of signs thus cover the four distinct regions delimited by the first order Bragg lines.

The evaluation of the second-order integral (\ref{eq:second_order}) {requires some caution} due to the delta function constraint with respect to the $\omega_1$ and $\omega_2$ variables, which has been analyzed in detail in \cite{lipa_RS86}. For a prescribed value of $\omega$, the admissible values of $\vect k_1$ follow closed contour lines in the $(k_{1x},k_{1y})$ plane. The contours have 2 disjoint components for $\omega<\sqrt 2\omega_B$ that merge into a single one  for $\omega>\sqrt 2\omega_B$. The intensification of the contour lines at the transition produces a singularity of the function $\sigma_2(\omega)$ at $\omega=\sqrt 2\omega_B$. A second singularity is caused by the EM kernel, that has a sharp maximum for $\scal{k_1}{k_2}=0$ (this would be a mathematical singularity without the finite impedance term). The locus of this singularity is a circle  which is found tangential to the contour line $\omega=2^{3/4}\simeq 1.68\ \omega_B$; this results in a secondary, sharp maximum at this frequency. Considering the finite integration time $T_i$ and the resulting finite frequency resolution $\Delta\omega=2\pi/T_i$, the theoretical Doppler spectrum is actually smeared by some function $\Phi(\omega)$ of width $\Delta\omega$ \cite{walsh_RS00}. This {allows} to replace the Dirac function $\delta(\omega-n_1\omega_1-n_2\omega_2)$ in the integrand of (\ref{eq:second_order}) by $\Phi(\omega-n_1\omega_1-n_2\omega_2)$
where the integration domain is now two-dimensional with no further constraint on the integration variable $\vect k_1$. This allows for a straightforward computation of $\sigma_2$ but requires a tight 2D sampling of the {\it k}-space that is numerically not efficient. An alternative computation can be obtained by reducing the two-dimensional integral (\ref{eq:second_order}) to a single one by eliminating the delta function through an appropriate change of variables. The standard way to do this \cite{lipa_RS86} is to use polar coordinates $(k_1,\theta)$ for the wave vector $\vect k_1$ and to operate a change a variable $(k_1,\theta)\to (\omega_1+\omega_2,\theta)$. However, the change of variable is only implicit and requires an extra numerical inversion for each value of $\omega$ to calculate the new integrand. {However, there exists a more natural change of variable that is more efficient and fully explicit. In spite of its simplicity, this alternative change of variable was introduced only a few years ago \cite{shahidi_OCEANS16,shahidi_OCEANS17,shahidi_JOE20}, for the purpose of improved nonlinear inversion of the wave spectrum. It applies only to deep water gravity waves, which follow the simple dispersion relation $\omega^2=gk$, a case that covers the vast majority of applications of HF radars. We reproduce in the following the main steps of the derivation using our own technique and notations.} To simplify the calculation we use {normalized} frequencies and wave numbers
\be
\begin{split}
  \kappa_1=\vect k_1/k_B,&\ \kappa_2=\vect k_2/k_B,\ \vect\kappa_B=\vect k_B/k_B,\\
  \nu_1=\omega_1/\omega_B&\ \nu_2=\omega_2/\omega_B\ \nu=\omega/\omega_B.
\end{split}
\ee
This is relevant because the kernel $\Gamma$ depends only on the {normalized} variables and the Bragg wavenumber
\be
\Gamma(\vect k_1,\vect k_2,\omega,\omega_B)=k_B\Gamma(\vect \kappa_1,\vect \kappa_2,\nu,1).
\ee
Assuming the {normalized} Bragg wave vector to be along the x-axis ($\vect \kappa_B=(1,0)$) we operate the change of variables
\be
(\kappa_{1x},\kappa_{1y})\to (\nu_1,\nu_2).
\ee
Note that we thus replace the two-dimensional wave vector $\kappa_1$ by a set of 2 positive frequencies with the obvious advantage that these appear as the argument of the delta function in the integrand.
\begin{figure*}[ht!]
  \begin{center}
    \includegraphics[scale=0.25]{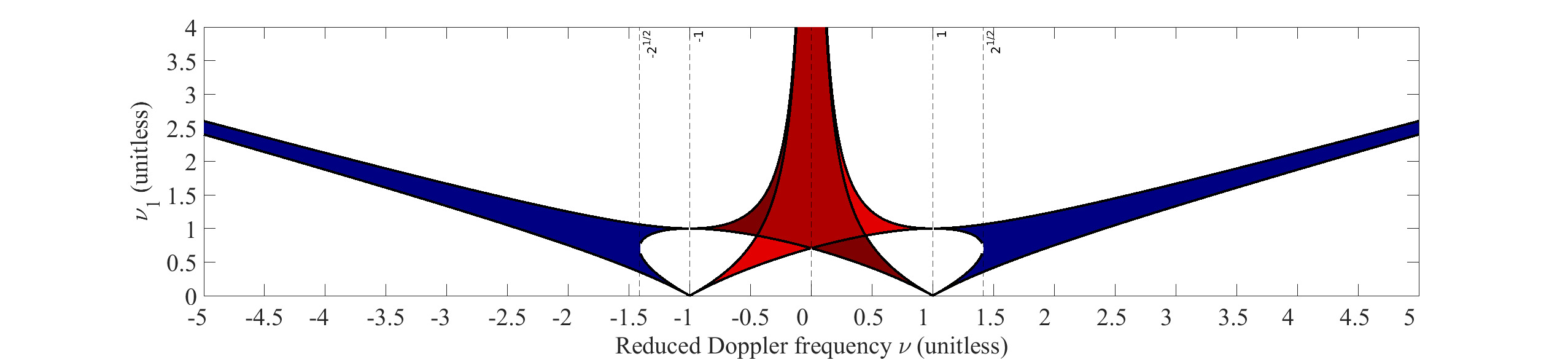}
         \caption{Admissible values of the {normalized} frequency $\nu_1$ seen in the $(\nu,\nu_1)$ plane. For $\abs\nu>1$, the integration domain of the integral (\ref{intomega1}) corresponds to vertical cuts of the blue strips.
       For $\abs\nu<1$, the integration domains of the integrals (\ref{intomega2int}) are given by vertical cuts of the dark red ($I_{-1,1}$) and light red  ($I_{1,-1}$) regions. \label{fig:domaineintegrationw1}}
\end{center}
\end{figure*}
On the upper half-space  $\kappa_{1y}\geq 0$ we can derive a simple one-to-one relationship between old and new integration variables
\be\label{eq:changevar}
\begin{split}
 \kappa_{1x}(\nu_1,\nu_2)&=\demi\left(1+\nu_1^4-\nu_2^4\right)\\
 \kappa_{1y}(\nu_1,\nu_2)&=\abs{\nu_1^4-\frac{1}{4}\left(1+\nu_1^4-\nu_2^4\right)^2}^{1/2}
\end{split}
\ee
and the same holds on the lower half-space by changing the sign of $\kappa_{1y}$ in  (\ref{eq:changevar}). The Jacobian of the transformation 
is found to be
\be
J(\nu_1,\nu_2)=\abs{\frac{4\nu_1^3\nu_2^3}{\kappa_{1y}(\nu_1,\nu_2)}}.
\ee
We now evaluate the second-order Doppler spectrum in this new set of variables by treating separately the frequency intervals between and outside the Bragg lines.

\paragraph{Case $\omega>\omega_B$ ($n_1=n_2=1$) or $\omega<-\omega_B$ ($n_1=n_2=-1$)}

By separating the integrals over the two half-spaces and performing the integration over the $\nu_2$ variable to eliminate the delta function we obtain
\be\label{intomega1}
\sigma_2(\omega)={\cali N} k_B^4\omega_B^{-1} \int_{I(\nu)} {\cali S}(\nu_1) \gamma(\nu_1) {\cali J}(\nu_1)\ d\nu_1
\ee
where $\gamma$ is the {normalized} kernel expressed as a function of {only} the {normalized} frequencies $\nu=\omega/\omega_B$ and $\nu_1$
\be\label{intomega1b}
\gamma(\nu_1)=\mc{\Gamma( \vect \kappa_1^+,\vect \kappa_2^+,\nu,1)}=\mc{\Gamma( \vect \kappa_1^-,\vect \kappa_2^-,\nu,1)},
\ee
$\vect \kappa_1^\pm,\vect \kappa_2^\pm$ are the {normalized} wave vectors in the upper/lower half-space
\be\label{intomega1c}
\begin{split}
\vect \kappa_1^\pm&=(\kappa_{1x}(\nu_1,\abs\nu-\nu_1), \pm {\kappa_{1y}(\nu_1,\abs\nu-\nu_1)})\\
\vect \kappa_2^\pm &=(1-\kappa_{1x}(\nu_1,\abs\nu-\nu_1), \mp{\kappa_{1y}(\nu_1,\abs\nu-\nu_1)}),
\end{split}
\ee
${\cali S}$ is the corresponding product of directional ocean spectra
\be\label{eq:S}
   {\cali S}(\nu_1)=S_d(n_1 k_B\vect \kappa_1^+)S_d(n_1 k_B\vect \kappa_2^+)+S_d(n_1 k_B\vect \kappa_1^-)S_d(n_1 k_B\vect \kappa_2^-)
   \ee
   and ${\cali J}$ is the Jacobian in the {normalized} frequency variable:
   \be\label{eq:jacobian}
            {\cali J}=J(\nu_1,\abs \nu-\nu_1).
      \ee
The integration domain $I(\nu)$ is a subset of the positive axis. It is limited to the values of $\nu_1$ for which $\nu_2\geq 0$ and for which the argument of the square root in (\ref{eq:changevar}) is positive, that is
\be
\nu_1^4-\frac{1}{4}\left(1+\nu_1^4-(\abs \nu-\nu_1)^4\right)^2\geq 0.
\ee
The limits of the domain can easily be identified analytically with the roots of the above equation. The integration domain is composed of one or two disjoint intervals depending on the value of $\nu$
\be\label{eq:Inu}
I(\nu)=\left\{\begin{array}{ll}
\left[\frac{\nu^2- 1}{2\abs\nu},\frac{\abs\nu-\sqrt{2-\nu^2}}{2}\right]\cup \left[\frac{\abs\nu+\sqrt{2-\nu^2}}{2}, \frac{1+\nu^2}{2\abs\nu}\right],&\ 1\leq \abs\nu \leq \sqrt 2,\\
\left[\frac{\nu^2-1}{2\abs\nu},\frac{\nu^2+ 1}{2\abs\nu}\right],&\ \abs\nu>\sqrt 2.
  \end{array}\right.
\ee

Note that the integration domain is {split} in two disjoints intervals for $1<\abs\nu<\sqrt 2$. However, these two intervals are interchanged by the transformation $\nu_1\mapsto \nu_1-\nu$. Since this transformation amounts to swap $\nu_1$ and $\nu_2$ in the integrand, this leaves  $\gamma$, $ {\cali J}$ and ${\cali S}$ unchanged. It follows that the 2 integrals on the lower and  upper interval are identical so that it suffices to calculate one of them and multiply the result by two.

\paragraph{Case $\abs\omega<\omega_B$ ($n_1=-n_2=\pm 1$)}

Similar calculations lead to the following expression for the second-order Doppler spectrum inside the Bragg lines:
\be\label{intomega2int}
\sigma_2(\omega)={\cali N} k_B^4\omega_B^{-1} \left\{\int_{I_{1,-1}(\nu)}+\int_{I_{-1,1}(\nu)}\right\} {\cali S}(\nu_1) \gamma(\nu_1) {\cali J}(\nu_1)\ d\nu_1
\ee
with the following adaptation of the integrand. On the interval $I_{1,-1}(\nu)$, the variable $\abs\nu-\nu_1$ should be replaced by $\nu_1-\nu$ in (\ref{intomega1c}) and (\ref{eq:jacobian}), and  (\ref{eq:S}) should be replaced by:
\be\label{eq:Spm}
   {\cali S}(\nu_1)=S_d(k_B\vect \kappa_1^+)S_d(-k_B\vect \kappa_2^+)+S_d(k_B\vect \kappa_1^-)S_d(-k_B\vect \kappa_2^-).
   \ee
   On the interval $I_{-1,1}(\nu)$, the variable $\abs\nu-\nu_1$ should be replaced by $\nu_1+\nu$ in (\ref{intomega1c}) and (\ref{eq:jacobian}), and  (\ref{eq:S}) should be replaced by
\be\label{eq:Smp}
   {\cali S}(\nu_1)=S_d(-k_B\vect \kappa_1^+)S_d(k_B\vect \kappa_2^+)+S_d(-k_B\vect \kappa_1^-)S_d(k_B\vect \kappa_2^-).
   \ee
The integration domain $I_{1,-1}(\nu)$ is constrained by the conditions
\be
\nu_1^4-\frac{1}{4}\left(1+\nu_1^4-(\nu-\nu_1)^4\right)^2\geq 0,\ \nu_1\geq \nu
\ee
leading to
\be
I_{1,-1}(\nu)=\left[\frac{\nu+\sqrt{2-\nu^2}}{2},\frac{1+\sign\nu\nu^2}{2\abs\nu}\right].
\ee
Likewise the integration domain $I_{-1,1}(\nu)$ is constrained by the conditions
\be
\nu_1^4-\frac{1}{4}\left(1+\nu_1^4-(\nu+\nu_1)^4\right)^2\geq 0,\ \nu_1\geq -\nu
\ee
leading to
\be
I_{-1,1}(\nu)=\left[\frac{-\nu+\sqrt{2-\nu^2}}{2},\frac{1-\sign\nu\nu^2}{2\abs\nu}\right].
\ee
The complete integration domain defined by the union of $I(\nu)$ (for $\abs\nu\geq 1$), $I_{1,-1}(\nu)$ and  $I_{-1,1}(\nu)$ (for $\abs\nu<1$)  is shown in the $(\nu,\nu_1)$ plane in Fig. \ref{fig:domaineintegrationw1}. Note that the transformation $\nu_1\mapsto \nu_1-\nu$ maps the interval  $I_{1,-1}(\nu)$ to $I_{-1,1}(\nu)$. Again, since this transformation amounts to swapping $\nu_1$ and $\nu_2$ in the integrand, this leaves  $\gamma$ and $ {\cali J}$ unchanged in the integral and transforms  (\ref{eq:Spm}) to  (\ref{eq:Smp}). It follows that the 2 integrals in  (\ref{intomega2int}) are identical so that it suffices to calculate one of them

\be\label{intomega2int2}
\sigma_2(\omega)=2{\cali N} k_B^4\omega_B^{-1} \int_{I_{-1,1}(\nu)}{\cali S}(\nu_1) \gamma(\nu_1) {\cali J}(\nu_1)\ d\nu_1.
\ee

When $\nu_1$ approaches the border of the domain, corresponding to one end of the intervals $I(\nu),I_{1,-1}(\nu)$ or $I_{-1,1}(\nu)$, the Jacobian term ${\cali J}(\nu_1)$ has an integrable $1/\sqrt \epsilon$ singularity, except at the particular values $\nu=\pm \sqrt 2$, where its denominator has a double root  causing a non-integrable $1/\epsilon$ singularity. This recovers the result established in \cite{ivonin_joe06} that the Doppler spectrum has a logarithmic singularity at $\nu=\pm \sqrt 2$. Apart from these 2 specific points, the singularity of the Jacobian at the borders of the domain can be handled properly by using the Gauss-Jabobi quadrature rule for numerical integration. We recall that
\be\label{eq:gaussjacobi}
\int_{-1}^{+1} (x-1)^{-1/2} (x+1)^{-1/2}f(x)dx\simeq \sum_1^N w_j f(x_j)
\ee
where the nodes of the quadrature $x_j$ are the zeros of the Jacobi polynomials and $w_j$ the associated weights, the approximation being exact whenever $f$ is constant. Many open source numerical routines can be found on the internet. This allows the calculation of the second-order ocean Doppler spectrum (\ref{intomega1}) at the cost of an ordinary one-dimensional integral.

\begin{figure*}[ht!]
  \begin{center}
    \includegraphics[scale=0.3]{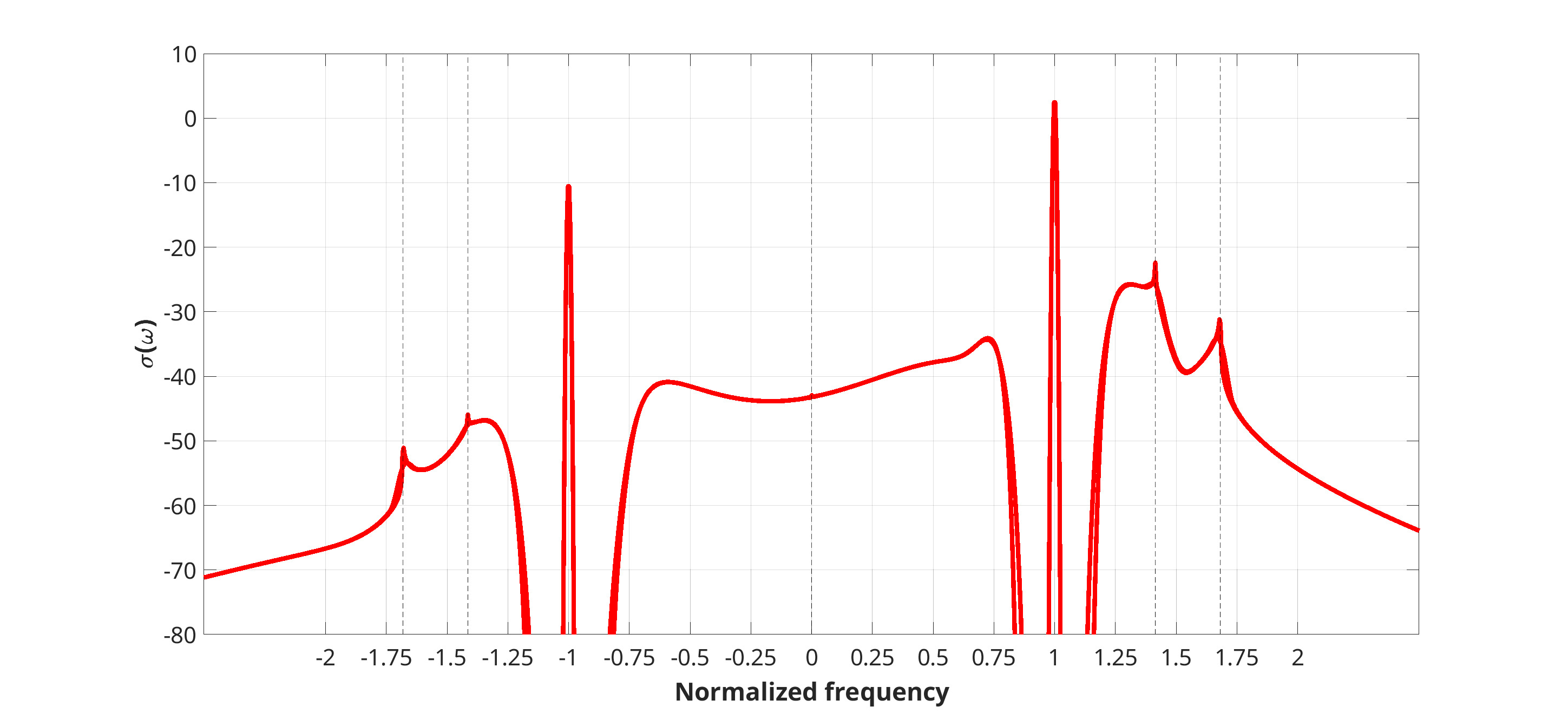}
           \caption{Second-order Doppler spectrum at radar frequency $f_0=16$ MHz calculated with the improved formulation (\ref{intomega1}) as a function of the {normalized} Doppler frequency $\nu=\omega/\omega_B$. The ocean wave spectrum is a directional PM spectrum at 10 m/s wind speed with the radar looking upwind (black solid lines). The Doppler spectrum has been smoothed according to a finite frequency resolution of $0.01$ Hz (red solid lines). The black dashed vertical lines mark the $\pm 2^{1/2}$ and $\pm 2^{3/4}$ singularities. \label{fig:simu2ordre}}
 \end{center}
\end{figure*}

The sea state dependence in the expression of the second-order Doppler spectrum enters through the ocean wave spectrum. Following the standard representation we write the directional wave spectrum in the form
\be
S_d(\vect k)=k^{-1}S_{o}(k)D(k,\theta)
\ee
where $S_{o}(k)$ is the omnidirectional spectrum and $D(k,\theta)$ is an angular spreading function with the normalization $\int_0^{2\pi}D(k,\theta)d\theta=1$ and the angle $\theta$ is measured with respect to the radar beam. It can also be expressed in term of the classical frequency spectrum $S_f(\omega)$ (using $S_{o}(k)=S_f(\omega)d\omega/dk$ and $\omega=\sqrt{gk}$)
\be\label{eq:relSdSf}
S_d(\vect k)=\demi g^2\omega^{-3}S_{f}(\omega)D(\omega,\theta).
\ee
For the numerical simulations we have employed the classical Pierson-Moskowitz (PM) frequency spectrum
\be
S_f(\omega)=Ag^2\omega^{-5}\exp(-B(g/U_{10}\omega)^4)
\ee
corresponding to an omnidirectional wavenumber spectrum
\be
S_{o}(k)=\frac{A}{2} k^{-3}\exp(-Bg^2/(U_{10}^4k^2))
\ee
where $A=0.0081$, $B=0.74$, $U_{10}$ is the wind speed at 10 meters above the sea surface. For the angular spreading function we chose the classical cardioïd shaped spreading function
\be
D(\theta)=\alpha(\epsilon+(1-\epsilon)\cos^4((\theta-\theta_w)/2))
\ee
where $\alpha$ is the normalization constant, $\theta_w$ is the direction of wind with respect to the radar beam ($\theta_w=0$ upwind) and $\epsilon=0.05$ is a small parameter that allows to have a fraction of waves propagating against the wind, as it is often observed. An example of the resulting computation for the upwind ocean Doppler spectrum at wind speed  $U_{10}=10$ m/s is shown in Fig. \ref{fig:simu2ordre} at the commonly used radar frequency $f_0=16$ MHz. We superimposed the smoothed version of this spectrum, which removes the delta-like singularity at $\omega=\pm\omega_B$ and the logarithmic singularity at $\omega=\pm\sqrt 2\omega_B$ and corresponds to a practical situation. The computation was assessed by comparison (not shown here) with the aforementioned direct method based on a two-dimensional integration in the $\vect k_1$ space using a finite filter $\Phi$ instead of a Dirac function.

\section{The Zeta function} \label{sec:zetafunction}
We now consider the following function, henceforth referred to as the ``Zeta function'', obtained by removing the coupling coefficient from the expression of the second-order ocean Doppler spectrum and replacing it with $k_B^2$ to keep the same dimension
\be\label{eq:zeta}
\begin{split}
  \zeta(\omega)&={\cali N} k_B^2 \sum_{n_1=\pm 1}\sum_{n_2=\pm 1} \int S_d(n_1\vect k_1)S_d(n_2\vect k_2)\\
  &\delta(\omega-n_1\omega_1-n_2\omega_2)d\vect k_1.
  \end{split}
\ee
As first suggested in \cite{barrick_RS77}, the main parameters of the ocean wave spectrum can be estimated with the first two moments of the Zeta function and we will follow the same argument. We first integrate the Zeta function over the frequency variable to get rid of the delta function in the integrand of $\zeta$. The second step of the demonstration is obtained by noting that the ocean wave spectrum has a dominant value near the peak wavenumber $k_p<<k_B$ so that its leading contribution can be extracted from the integral. This occurs when one of the variable $\vect k_1$ or $\vect k_2$ in the integrand is close to zero, leading in the end to
\be\label{approxintzeta}
\int_{-\infty}^{+\infty} \zeta(\omega)d\omega\simeq 4{\cali N} k_B^2 \sum_{n_1=\pm 1} S_d(n_1\vect k_B)\int S_d(\vect k) d\vect k.
 \ee
 The integral of the right-hand side is the variance of wave elevations
 \be\label{H0zeta}
 H_0^2=\int S_d(\vect k) d\vect k
 \ee
 while the discrete summation is the integral of the first-order Bragg peaks. Hence the Significant Wave Height (SWH) $H_s=4H_0$ can be recovered from the Zeta function
 \be
 H_s^2= \frac{4\int_{-\infty}^{+\infty} \zeta(\omega)d\omega}{k_B^2\int_{-\infty}^{+\infty} \sigma_1(\omega)d\omega}.
 \ee
To estimate the mean wave frequency we consider the positive frequency side of the Zeta function spectrum (assuming it is stronger than the negative frequency side) and take the following ratio
    \be\label{Tzeta}
    {\Omega}=\frac{\int_{\omega_B}^{+\infty} (\omega-\omega_B)\zeta(\omega)d\omega}{ \int_{\omega_B}^{+\infty} \zeta(\omega)d\omega}.
    \ee
After integration of the delta function this can be rewritten as
    \be
    {\Omega}= \frac{\int (\omega_1+\omega_2- \omega_B)S_d(\vect k_1)S_d(\vect k_2) d\vect k_1}{\int S_d(\vect k_1)S_d(\vect k_2) d\vect k_1}.
    \ee
 Again, the leading contribution of the integrand occurs when one of the frequency $k_1$ or $k_2$ is close to zero and the other close to $k_B$, so that one of the $\omega_1,\omega_2$ variable cancel out with the $-\omega_B$ term. This identifies $\Omega$ as the centroïd frequency
    \be\label{eq:Omega}
    {\Omega}\simeq \frac{\int \omega(\vect k)S_d(\vect k)d\vect k}{\int S_d(\vect k)d\vect k}=\frac{2\pi}{T}
    \ee
which defines the mean period $T$.
 
 \begin{figure*}[ht!]
    \hspace{-1cm} \includegraphics[scale=0.35]{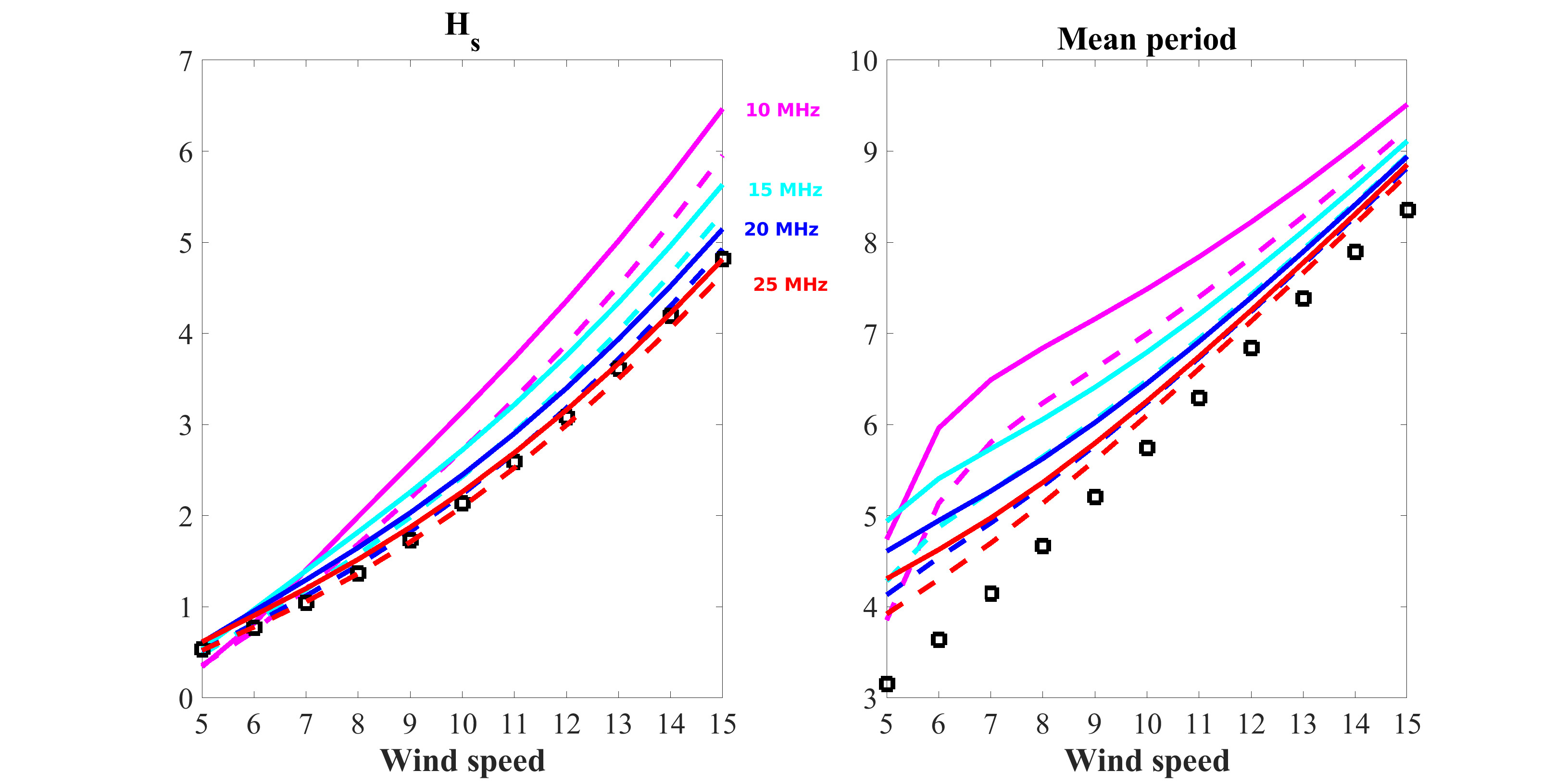}
     \caption{Estimation of a) the SWH and b) the mean period from the Zeta function as a function of wind speed for a directional PM spectrum with the radar aiming in the upwind (solid lines) or crosswind (dashed lines) direction. The black squares show the theoretical value of the SWH and mean period. The various radar frequencies correspond to different color codes: 10 MHz (magenta); 15 MHz (cyan); 20 MHz (blue) ; 25 MHz (red). \label{fig:evolHs_f0_upcross}}
       \end{figure*}

We performed numerical simulations, shown in  Fig. \ref{fig:evolHs_f0_upcross}, to estimate the SWH and the mean wave period $T$ from the Zeta function for the directional PM spectrum at various wind speed and direction at several radar frequencies covering the typical range of HF radars (from 10 to 25 MHz). We found that the SWH is increasingly well estimated by formula (\ref{H0zeta}) as the radar frequency varies from 10 to 25 MHz but is poorly estimated at the lowest frequency (10 MHz). The same conclusion applies to the mean period, which is overestimated by formula (\ref{Tzeta}-\ref{eq:Omega}) but becomes more accurate at higher radar frequencies and larger wind speed. This is consistent with the expectation that the approximation that was used to evaluate the integral of the Zeta function (\ref{approxintzeta}) necessitates a high ratio of the Bragg wavenumber $k_B$ to the peak wavenumber $k_p$. The observed dependence of the estimated wave parameters on the radar frequency in Fig. (\ref{fig:evolHs_f0_upcross}) suggests that the estimated SWH ($\hat H_s$) should be corrected for by a multiplicative bias while the estimated mean period ($\hat T$) should be corrected for by an additive bias
       \be \label{eq:bias1}
       \hat H_s\to \alpha \hat H_s,\ \hat T\to \hat T-T_0.
       \ee
By adjusting the average of the upwind and cross-wind estimation to the theoretical wave parameters for the ensemble of wind speeds we obtained parameters $\alpha$ and $T_0$ that depend on the sole radar frequency. These values are summarized in Table \ref{table:alphaT0}. The corrective factor for the SWH ranges from $0.93$ to $0.97$ when increasing the radar frequency to $10$ to $25$ MHz (we discard the case $5$ MHz) while the bias to subtract from the mean period ranges from $0.4$ to $1.25$ seconds. 
       \begin{table}
         \begin{center} \caption{Corrective factors to apply in the derivation of the sea state parameters from the Zeta function\label{table:alphaT0}}
  \begin{tabular}{|l|l|l|l|l|}
    \hline
    Radar frequency (MHz) & 10 & 15 & 20 & 25 \\
    \hline
    $\alpha$ &  0.93&    0.95&    0.96&    0.97\\
    \hline
    $T_0$ (sec) & 1.25&    0.76&    0.53&    0.4\\
    \hline
  \end{tabular}
  \end{center}\label{eq:bias2}
   \end{table}

       \section{The weighting function}\label{sec:wf}
\begin{figure}[ht!]
\hspace{-0.5cm}  \includegraphics[scale=0.2]{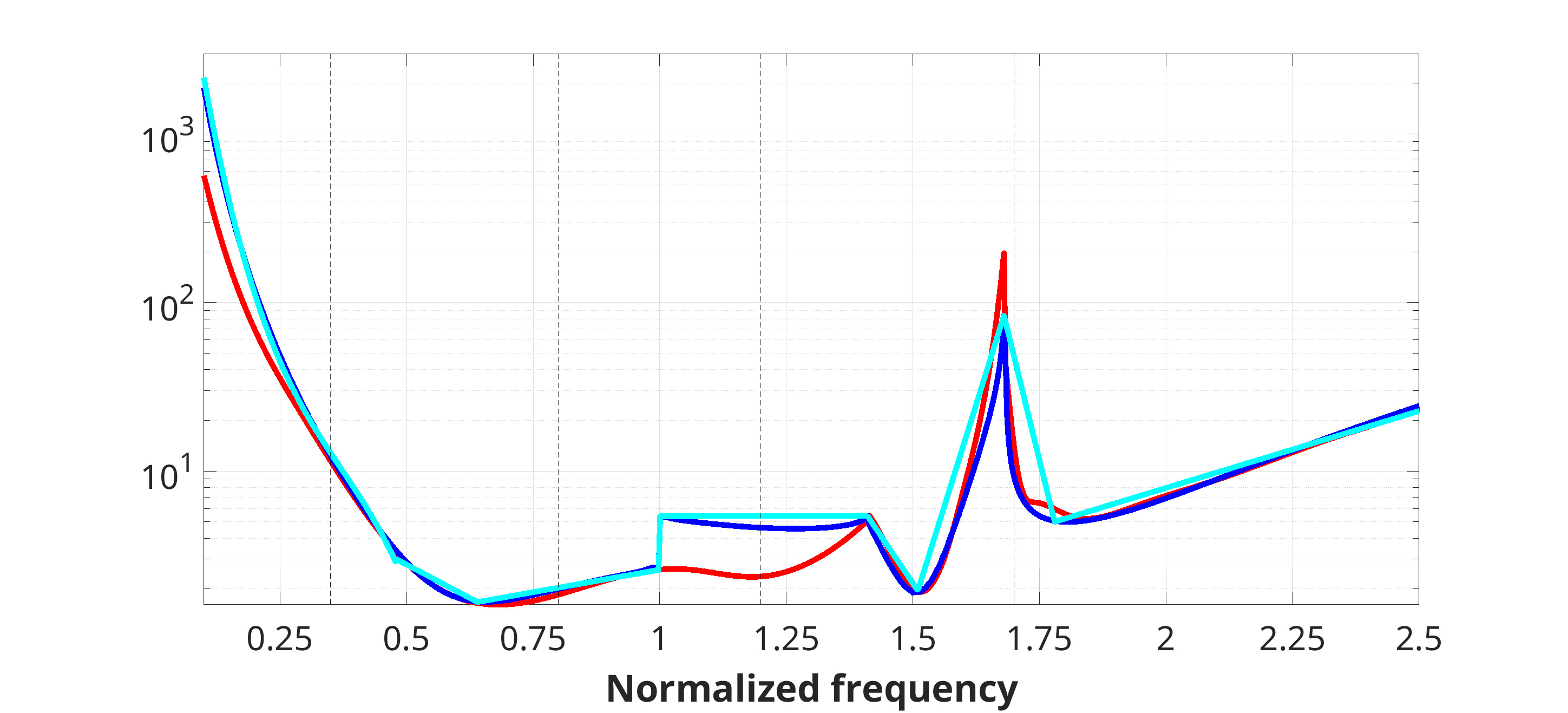}
      \caption{New weighting function (blue solid line) as compared to Barrick's weighting function (red solid line). The simple analytical fit (\ref{eq:fitWF}) is also shown (cyan line). The dashed vertical lines mark the 2 approximate frequency intervals ($[0.35-0.8]$ and $[1.2-1.7]$ where the weighting function is useful.\label{fig:wf}}
\end{figure}
\begin{figure*}[ht!]
   \hspace{-0cm}\includegraphics[scale=0.34]{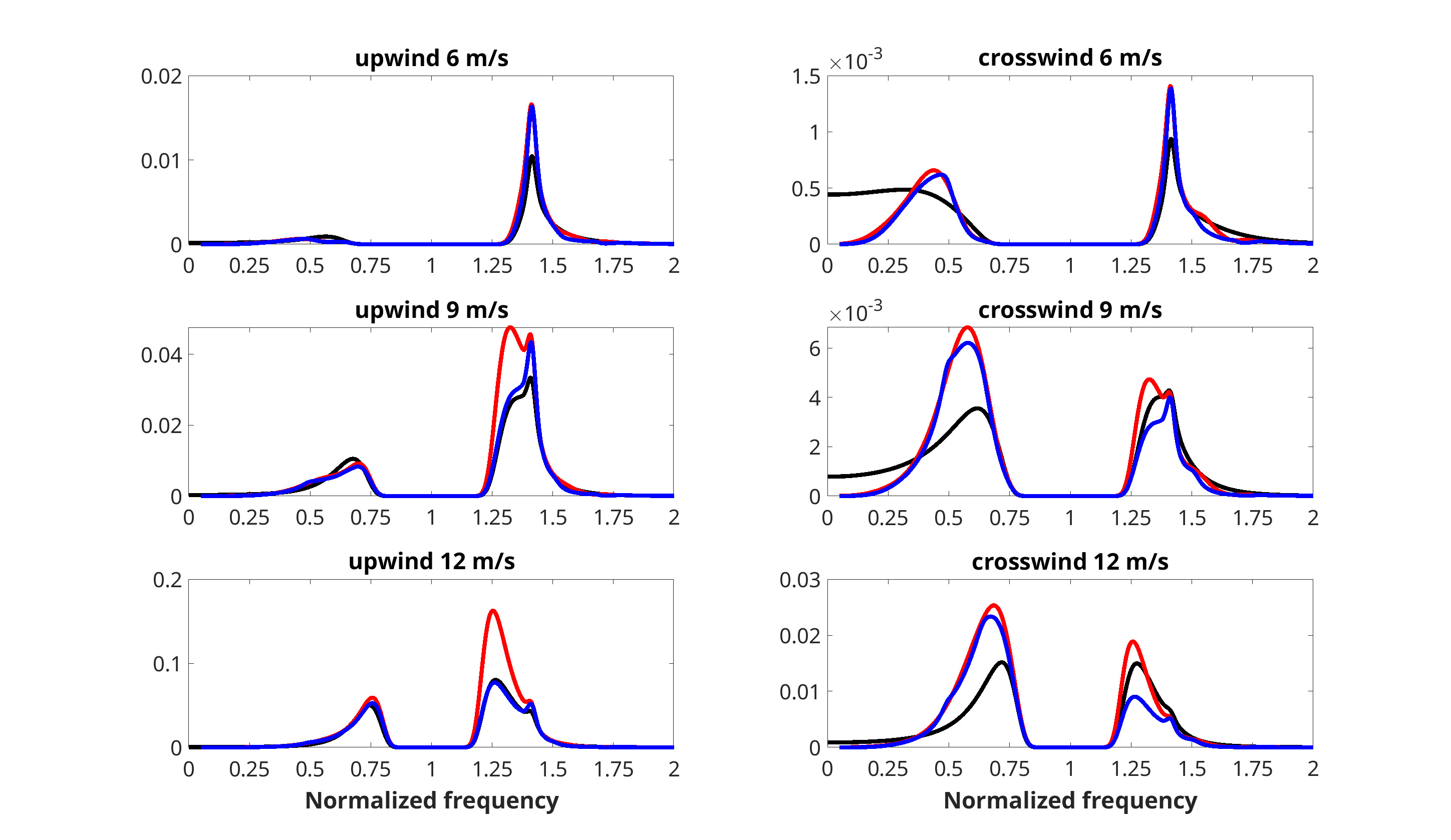}
     \caption{Comparison of the Zeta function (black lines) and its evaluation with the ocean Doppler spectrum ( \ref{eq:approxsigma2}) using Barrick's weighting function (red line) and ours with the simple fit (\ref{eq:fitWF}) (blue lines). The ocean Doppler spectrum was simulated  at the radar frequency $16$ MHz using \col{ a directional PM spectrum in upwind and crosswind direction at wind speed 6, 9 and 12 m/s}. The three functions have been smoothed with a window of width 0.01 Hz and are shown in linear scale (i.e., not dB) to highlight the frequency intervals where they are non-negligible. \label{fig:compaapproxwf}}
\end{figure*}

In order to make the expression of the ocean Doppler spectrum analytically tractable for sea state inversion, Barrick \cite{barrick_RSE77} proposed to replace the coupling coefficient $\mc\Gamma$ by the so-called weighting function, obtained  by performing an angular average $\overline{\mc\Gamma}$ over the admissible wavenumbers
\be
\frac{k_0^2}{8}{\cali W_B}(\nu)=\overline{\mc\Gamma}.
\ee
We found this weighting function difficult to implement from the original publication and we used the numerical routine provided by the free RadarWIC library \cite{cahl_WaveRIC19}, that uses a digitized version of the Fig. 3 of \cite{barrick_RSE77}. The reformulation of the second-order integral (\ref{intomega1}) suggests the construction of a different weighting function, using the same normalization factor as Barrick
\be\label{defwfB}
\frac{k_0^2}{8}{\cali W}(\nu)=k_B^2\langle\gamma\rangle_{I(\nu)}
\ee
where $\langle{\gamma}\rangle_{I(\nu)}$ is an average value of $\gamma$ over the {normalized} frequency domain $I_\nu$. For $\nu>1$, this domain is narrow and it is justified to defined it with the mean value of the coupling coefficient over this interval
\be\label{defwf}
\langle\gamma\rangle_{I(\nu)}=\frac{\int_{I(\nu)}\gamma(\nu_1)d\nu_1}{\int_{I(\nu)}d\nu_1},\ \nu>1.
\ee
\col{For $1<\nu<\sqrt 2$}, the values of both the kernel $\gamma$ and the product of spectra in the integrands (\ref{intomega2int}) are more contrasted, so that is not relevant to retain the mean value. However, we observed numerically and checked analytically that the integrand reach their dominant value at the edges of the integration domain. To give more weight to the latter, we replaced the mean by a simple 3 points average between the ends and the middle of the lower interval $I_{-1,1}(\nu)$ (we recall that the integral on the upper interval $I_{1,-1}(\nu)$ is identical)
\be\label{defwf2}
\langle\gamma\rangle_{I(\nu)}=\frac{1}{3}\left(\gamma(A_\nu)+\gamma\left(\frac{A_\nu+B_\nu}{2}\right)+\gamma(B_\nu)\right)
\ee
with
\be\label{eq:meangamma2}
A(\nu)=\frac{\nu^2- 1}{2\nu},\ B(\nu)=\frac{\nu-\sqrt{2-\nu^2}}{2}.
\ee
For $\nu<0$, the weighting function is extended by parity (i.e., ${\cali W}(-\nu)={\cali W}(\nu)$). Barrick's weighting function and ours are shown in Fig. \ref{fig:wf}. Both exhibit a singularity at $\nu=2^{3/4}$ and follow the same trends, \col{with some differences, however, due to choice of integration variable and to the type of average.} The new weighting function can be easily implemented from (\ref{defwf}) and (\ref{eq:meangamma2}) but it is more convenient to have an analytical expression in the form of an empirical fit. We found the following elementary linear fit, which is enough for our purposes:
\be\label{eq:fitWF}
   {\cali W}(\nu)=\left\{
   \begin{split}
     \expo^{-129.1\nu^3+143.4\nu^2-63.4\nu+12.7}&\ \mathrm{if}\ \nu<0.48\\
    3\ \expo^{-3.66 (\nu - 0.48)} &\ \mathrm{if}\  0.48 < \nu < 0.64\\
    1.67\ \expo^{1.23 (\nu - 0.64)} &\ \mathrm{if}\  0.64 < \nu < 1\\
    5.43\ & \mathrm{if}\  1 < \nu < 2^{1/2}\\
    5.43\ \expo^{-10.24 (\nu - 1.41)} &\ \mathrm{if}\  2^{1/2} < \nu < 1.51\\
    1.95\ \expo^{22.11 (\nu - 1.51)} &\ \mathrm{if}\  1.51 < \nu < 2^{3/4}\\
    83.6\ \expo^{-28.17 (\nu - 1.68)} &\ \mathrm{if}\ 2^{3/4} < \nu < 1.78\\
    5\ \expo^{2.1 (\nu - 1.78)} &\ \mathrm{if}\  \nu>1.78 
      \end{split}
   \right.
   \ee
  Note that the new weighting function has a discontinuity at $\nu=1$, due to the adaptation of the definition of the mean kernel for $0<\nu<1$. However, this discontinuity has no consequence since the weighting function is only useful in two small intervals of frequencies lying on each side of the Bragg frequency, which are approximately $[0.35-0.8]$ and $[1.2-1.7]$. They correspond to the intervals where the Zeta function has non-negligible values (see Fig. \ref{fig:compaapproxwf}).
   
\begin{figure*}[ht!]
   \centering \includegraphics[scale=0.325]{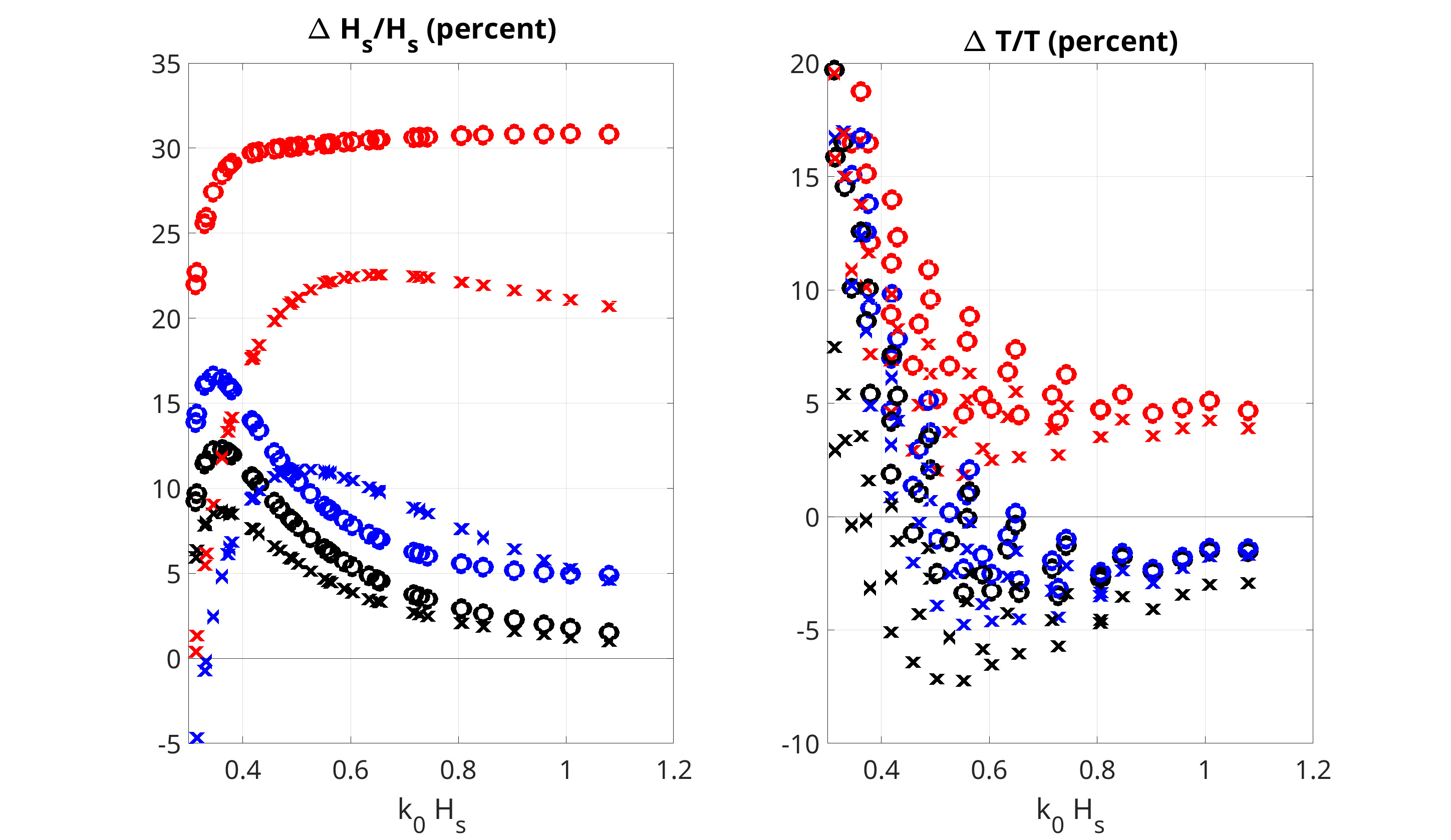}   
       \caption{Relative error (in percent) in the estimation of a) the SWH and b) the mean period from the second-order Doppler spectrum (eqs. (\ref{H0sigma2})-(\ref{Tsigma2})) using either Barrick's weighting function (red symbols) or fitted weighting function (\ref{eq:fitWF}) (blue symbols). The errors are shown as a function of the dimensionless roughness parameter $k_0H_s$, wherein the results of the four frequency bands 10,15,20 and 25 MHz have been concatenated.  The estimation based on the Zeta function is shown for reference in black symbols. The circles correspond to the upwind direction and the crosses to the crosswind direction.\label{fig:DeltaHs}}
\end{figure*}

Approximating the coupling coefficient with its average value in the expression (\ref{intomega1}) of the second-order Doppler spectrum leads to the following relation with the Zeta function
\be\label{eq:approxsigma2}
\zeta(\omega)\simeq 2^5\frac{\sigma_2(\omega)}{{\cali W}(\omega/\omega_B)}.
\ee
The properties of the Zeta function therefore transfer to the ocean Doppler spectrum provided the weighting function approximation is sufficiently accurate. In particular this leads to the following formulas for the estimation of the SWH and mean period
\be\label{H0sigma2}
(k_0H_0)^2=2\alpha^2\int_{-\infty}^{+\infty} R_{{\cali W}}(\omega)d\omega
\ee
and
\be\label{Tsigma2}
T=\frac{2\pi\int_{\omega_B}^{+\infty} R_{{\cali W}}(\omega)d\omega}{\int_{\omega_B}^{+\infty} (\omega-\omega_B)R_{{\cali W}}(\omega)d\omega}-T_0
\ee
where we define the ratio of the weighted second-order to first order Doppler spectrum following the notation of \cite{alattabi_JTECH19} as
\be
R_{{\cali W}}(\omega)=\frac{\sigma_2(\omega)/{\cali W}(\omega/\omega_B)}{\int_{-\infty}^{+\infty} \sigma_1(\omega)d\omega}.
 \ee
Apart from the corrective factors $\alpha$ and $T_0$ these expressions were already established by Barrick ( (11) and (13) of \cite{barrick_RS77}). 
Their accuracy is related, on one hand on the ability of the weighting function to recover the Zeta function (\ref{eq:approxsigma2}), and on the other hand on the quality of the estimation (\ref{H0zeta}), (\ref{Tzeta}) and (\ref{eq:bias1}) based on the Zeta function. To elucidate these points, we performed numerical simulations at various wind speed and radar frequencies using the directional PM spectrum.
Fig. \ref{fig:compaapproxwf} shows a comparison between the exact expression of the Zeta function and its approximation (\ref{eq:approxsigma2}) at the radar frequency 16 MHz. We tested both Barrick's weighting function and the fitted version of the new weighting function (\ref{eq:fitWF}). From a visual inspection (which is confirmed numerically) it is clear than the latter is in general more accurate, especially in the upwind direction. As a result, the estimation of the sea state parameters from the second-order Doppler spectrum (eqs. (\ref{H0sigma2})-(\ref{Tsigma2})) is improved when using this new weighting function, even in its elementary fitted version (\ref{eq:fitWF}). Fig. \ref{fig:DeltaHs} shows the accuracy of the estimation of the SWH and the mean period with the various weighting functions. The percentage of error between the estimated and actual values of the SWH and mean period as a function of dimensionless parameter $k_0H_s$ for a directional PM wave spectrum at various wind speeds and radar frequencies. For reference, the result of the estimation with the Zeta function  (eqs (\ref{H0zeta}), (\ref{Tzeta}) and (\ref{eq:bias1})) is also shown (\col{with no bias correction}). \col{As already noted by other authors (e.g. \cite{wyatt_book21}), inconsistent estimations are obtained in the upwind and crosswind direction using Barrick's weighting function; the estimated value of SWH is overestimated by $20-30\%$ and that of the mean period by $5-15\%$.  An overall error reduction is obtained with the new weighting function for $k_0H_s>0.5$ as well as a reduction of the upwind/crosswind discrepancy. As seen, there is also a small discrepancy between the upwind and crosswind estimation from the Zeta function. This sets up a limit to what could be obtained with an optimal weighting function.}

Note that an empirical correction (\ref{H0sigma2}) for the SWH arising from Barrick's formula has already been employed by several authors to improve its consistency with High-Frequency radar measurements \cite{heron_JAOT98,maresca_JGR80,ramos_JAOT09,morales_coasteng25}. At a radar frequency of 25.4 MHz, \cite{ramos_JAOT09} found the best fit with $\alpha\simeq 0.58$ and \cite{heron_JAOT98} with $\alpha\simeq 0.55$ {while \cite{morales_coasteng25} found $\alpha\simeq 0.58$ at 16.15 MHz.} \col{Based on our simulations with a directional Pierson-Moskowitz model (Fig. \ref{fig:compaapproxwf}), the corrective factor would be $\alpha\simeq 0.75$ or $\alpha\simeq 0.82$ depending on the wind direction.}

 \section{A direct relation to the wave spectrum}\label{sec:directinversion}
\begin{figure}[ht!]
          \includegraphics[scale=0.17]{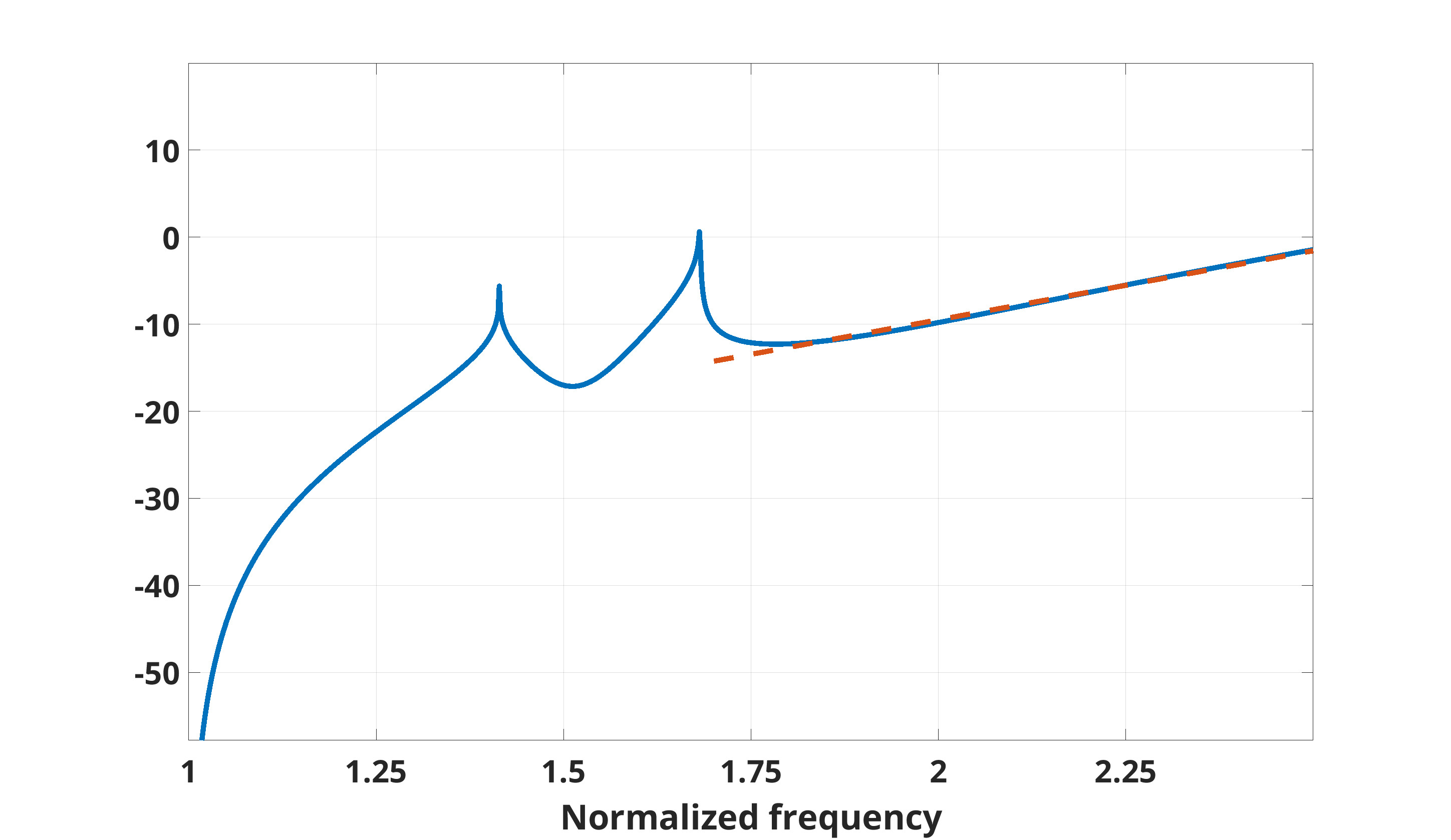}
                    \caption{Function ${\cali F}(\nu)$ (blue solid line) as defined in (\ref{eq:defF}) and expressed in decibels, together with its fit (\ref{fitF}) for $\nu>1.7$ (dashed red line).\label{fig:fonctionF}}
\end{figure}

\begin{figure*}[ht!]
    \centering
\includegraphics[scale=0.25]{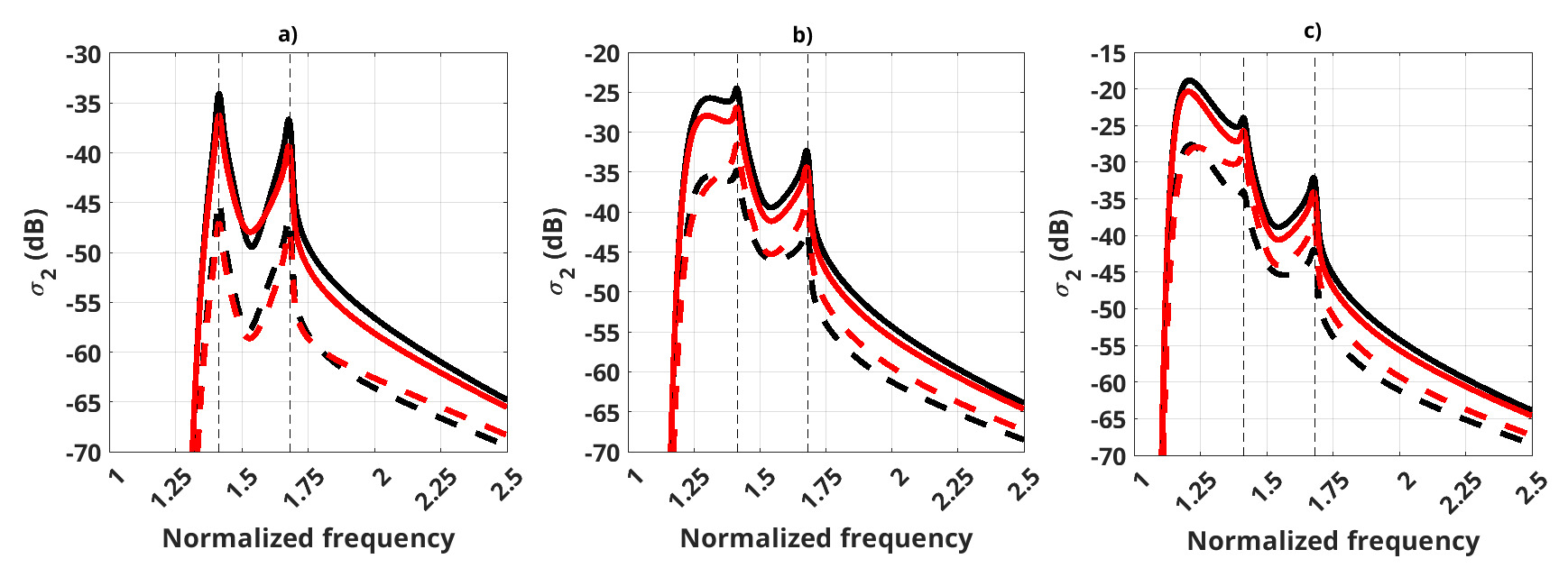}
  \caption{Exact numerical calculation (black lines) of the second-order Doppler spectrum compared with its analytical approximation with s (\ref{intomega_approx}) (red lines) for a directional PM spectrum. The radar is looking upwind (solid lines) or crosswind (dashed lines) at 16 MHz and the wind speed is a) 7 m/s; b) 10 m/s; c) 15 m/s.\label{fig:example_approxspectrum}}\end{figure*}

 In a series of papers \cite{gurgel_JOE06,lopez_JTECH16,alattabi_JTECH19,alattabi_JTECH21}, attempts were made to find a simple linear relationship between the ocean wave spectrum and the second-order Doppler spectrum shift by the Bragg frequency, as originally proposed by \cite{hasselmann_Nature71}. An empirical relation of the type
 \be
k_0^2 S_f(\omega)=2 A(\omega)R_{{\cali W}}(\omega\pm \omega_B)
 \ee
 which can be seen as the spectral form of Barrick's formula (\ref{H0sigma2}), has been sought with a frequency-dependent calibration coefficient $A(\omega)$ and different weighting functions (taken to be one for \cite{gurgel_JOE06,lopez_JTECH16} and adapted to wind-waves and swell for \cite{alattabi_JTECH19,alattabi_JTECH21}). Even though this method has shown good results when employed with experimental data, it is not fully satisfactory as it requires manual calibration for each radar site and bearing and is limited to the range of frequencies where the second-order Doppler spectrum can be linearized, i.e. not too far from the Bragg frequency. We propose here a nonlinear approximation, inspired by the reformulation (\ref{intomega1}) of the second-order Doppler spectrum as a frequency integral over the variable domain $I(\nu)$. For $\abs\nu>1$, the integration domain $I(\nu)$ is narrow (see Fig. \ref{fig:domaineintegrationw1}) and we can therefore approximate the above integral by extracting the mean value $\overline{{\cali S}}$ of the function ${\cali S}$ over this interval, leading to
\be\label{intomega_approx}
\sigma_2(\omega)\simeq {\cali N} k_B^4\omega_B^{-1} \overline{{\cali S}}(\nu) {\cali F}(\nu)
\ee
where
\be\label{eq:defF}
   {\cali F}(\nu)=\int_{I(\nu)} \gamma(\nu_1){\cali J}(\nu_1)\ d\nu_1
   \ee
is a deterministic function that does not depend on sea state nor on the radar frequency and can thus be calculated once for all. This function, which is shown in Fig. \ref{fig:fonctionF}, carries the 2 singularities at $\nu=2^{1/2}$ and $\nu=2^{3/4}$ but has otherwise a smooth polynomial behavior. For $\nu>1.7$ it can be described by the following fit with an excellent accuracy
   \be\label{fitF}
      10\log_{10}{\cali F}(\nu)=15.865\nu-41.23
      \ee
To obtain a tractable expression we define $\overline{{\cali S}}$ in the following way. For $1\leq \nu \leq \sqrt 2$, the integration domain  $I(\nu)$ consists of 2 narrow intervals given by (\ref{eq:Inu}) and we may approximate ${\cali S}$ by its  average between the endpoints of each interval; for $\nu>\sqrt 2$, the integration domain  $I(\nu)$ consists of a single wider interval on which we may approximate ${\cali S}$ by its  average between the midpoint and the endpoints. Altogether, this leads to the following definition
\be\label{meanS1}
\overline{{\cali S}}(\nu)=\left\{
\begin{array}{l}
  \frac{1}{2}({\cali S}_1+{\cali S}_2),\ \mathrm{if}\ 1\leq \nu\leq \sqrt 2\\
  \\
  \frac{1}{2}({\cali S}_0+{\cali S}_2),\  \mathrm{if}\ \nu>\sqrt 2
\end{array}
\right.
\ee
with
\be
{\cali S}_0= {\cali S}(\nu/2),\ {\cali S}_1= {\cali S}\left(\frac{\nu\pm \sqrt{2-\nu^2}}{2}\right),\ {\cali S}_2={\cali S}\left(\frac{\nu^2\pm 1}{2\nu}\right).
   \ee
   From the relation (\ref{eq:relSdSf}) we can convert this last expression in terms of the directional frequency spectrum
\be\label{eq:Snu}
\begin{split}
  {\cali S}_0&=\frac{32 g^4}{\omega^{6}} S^2_f(\omega/2)(D(\omega/2,\theta_0)D(\omega/2,-\theta_0)),\\
  {\cali S}_1 &=\demi\frac{g^4}{(\omega_1^+\omega_1^-)^3} S_f(\omega^-_1)S_f(\omega^+_1)D(\omega^+_1,0)D(\omega^-_1,0)\\
  {\cali S}_2 &= \demi\frac{g^4}{(\omega_2^+\omega_2^-)^3} S_f(\omega^-_2)S_f(\omega^+_2)D(\omega^+_2,0)D(\omega^-_2,0)\\
\omega_1^\pm&=\frac{\omega\pm\sqrt{2\omega_B^2-\omega^2}}{2},\ \omega^\pm_2=\frac{\omega^2\pm\omega^2_B}{2\omega}\\
\theta_0&=\mathrm{atan}\left(\demi\sqrt{\frac{\omega^4}{\omega_B^4}-4}\right).
\end{split}
  \ee
  Similar relations hold for negative frequencies ($\nu<-1$), with the argument of the spreading function shifted by $\pi$ ($D(\omega/2,\pm\theta_0)\to D(\abs\omega/2,\pm\theta_0+\pi)$, $D(\omega_{1,2}^\pm,0)\to D(\abs{\omega_{1,2}^\pm},\pi)$). Fig. \ref{fig:example_approxspectrum} shows a comparison of the exact formula (\ref{intomega1}) and its approximated version using (\ref{intomega_approx}) for a PM spectrum by 10 m/s wind speed for an upwind looking radar at 16 MHz. We found an overall agreement better than 2 dB. We checked numerically other wind speeds and radar frequencies with the same conclusion. This formula shows that starting from $\omega=2^{1/2}\omega_B$, the part of the ocean wave frequency spectrum that contributes to the Doppler spectrum at a given Doppler frequency $\omega$ is a narrow interval around $\omega/2$. The directions of waves involved in this process are those having an angle smaller than $\theta_0$ with respect to the radar beam. This last angle is null at $\omega=2^{1/2}\omega_B$, reaches $45$ degree at $2^{3/4}\omega_B$, and  $60$ degree at $2\omega_B$ and converges to the cross-beam directions $90$ degree as the frequency keeps increasing. Note that in the vicinity of the first singularity $2^{1/2}\omega_B$, the angle $\theta_0$ is close to zero so that the radar dominantly ``sees'' the waves that are propagating along or against the radar look direction. {This explains why the wave inversion with a single radar station gives only poor estimates of the crossbeam wave directions. It is interesting to evaluate the formulae (\ref{eq:Snu}) at the specific normalized frequency $\nu=\sqrt 2$, for which we obtain easily
 \be
 \frac{{\cali S}_0}{{\cali S}_2}=\frac{2^3}{3^3}  \frac{S_f(3\omega_B/2)S_f(\omega_B/2)}{S^2_f(\omega_B/\sqrt 2)} \frac{D(3\omega_B/2,0)D(\omega_B/2,0)}{D^2(\omega_B/\sqrt 2,0)} 
 \ee
For a Pierson-Moskowitz spectrum with $\omega^{-5}$ decrease, this yields the estimate
\be
\frac{{\cali S}_0}{{\cali S}_2}\simeq \frac{2^8}{3^8}\simeq 0.04
\ee
so that $\overline{{\cali S}}(\nu)\simeq \demi {\cali S}_0$
It follows from (\ref{intomega_approx}) and (\ref{meanS1}) that the integrated Doppler spectrum in the vicinity of the frequency $\sqrt 2\omega_B$ is  proportional to the wave frequency spectrum at the frequency $\omega_B/\sqrt 2$ in the upbeam direction ($\theta_0=0$)
\be
\int_{\sqrt 2\omega_B-\delta\omega}^{\sqrt 2\omega_B+\delta\omega}\sigma_2(\omega)d\omega \sim \omega_B S^2_f(\omega_B/\sqrt 2)D^2(\omega_B/\sqrt 2,0)\delta\omega,
\ee
where the integration over a small frequency interval of width $2\delta\omega<<\omega_B$ is necessary to remove the singularity of the function ${\cali F}(\nu)$ at $\sqrt 2$. The same holds around the negative Doppler frequency  $-\sqrt 2\omega_B$, which is related to the frequency wave spectrum in the downbeam direction ($\theta_0=\pi$). As a result, the upbeam/downbeam ratio for the angular spreading function at the frequency $\sqrt 2\omega_B$ can be obtained through:
\be
\frac{\int_{\sqrt 2\omega_B-\delta\omega}^{\sqrt 2\omega_B+\delta\omega}\sigma_2(\omega)d\omega}{\int_{-\sqrt 2\omega_B-\delta\omega}^{-\sqrt 2\omega_B+\delta\omega}\sigma_2(\omega)d\omega}=\frac{D^2(\omega_B/\sqrt 2,0)}{D^2(\omega_B/\sqrt 2,\pi)}.
\ee
This is a useful complement to the classical determination of the spreading function through the first-order Bragg peaks:

\be
\frac{\int_{\omega_B-\delta\omega}^{\omega_B+\delta\omega}\sigma_1(\omega)d\omega}{\int_{-\omega_B-\delta\omega}^{-\omega_B+\delta\omega}\sigma_1(\omega)d\omega}=\frac{D(\omega_B,0)}{D(\omega_B,\pi)}
\ee

}
  \section{Conclusion}
  We have {used a recently proposed} formulation of the second-order ocean Doppler spectrum based on an integration of the frequency variable {only}. This {alternative} expression allows for a more efficient numerical evaluation of the Doppler spectrum. In addition, it leads naturally to an alternative weighting function that can be used to improve the estimation of the main sea state parameters after the classical method proposed by Barrick in 1977 (\cite{barrick_RS77,barrick_RSE77}) and reduces the discrepancy between the upwind and crosswind estimation. Nevertheless, there are some limitations to this method, which are not related to the quality of the weighting function but to the structure of the Zeta function. A bias correction for both the SWH and mean period is necessary, with values depending on the radar frequency, and there remains a small but non-negligible discrepancy between the upwind and crosswind estimation from a single radar site. A second important consequence of our approach is the derivation of an analytical nonlinear approximation for the ocean wave spectrum for Doppler frequencies well beyond the Bragg frequency. This formula might be a useful complement to classical techniques of ocean wave spectrum inversion which are often limited to the frequencies close to the Bragg frequency (typically in the range $0.4<\nu< 1.6$) where the problem can be linearized (see e.g. the discussions in \cite{wyatt_JAOT00} or \cite{wyatt_book21}).

\begin{center}Acknowledgment\end{center}
  
{\small This research has been funded by the Agence Nationale de la Recherche (ANR) under grant ANR-22-ASTR-0006-01 (ROSMED project: ``Radar à Ondes de Surface en MEDiterranée'')}       


\end{document}